\documentclass[prd,aps,nofootinbib,preprintnumbers]{revtex4}
\usepackage{amssymb}
\usepackage{bm}
\usepackage{enumerate}
\usepackage{amssymb}
\usepackage{amsmath}
\usepackage{feynmf}
\usepackage{slashed}
\usepackage{wrapfig}
\usepackage{subfig}
\usepackage{graphicx}
\usepackage{filecontents}
\usepackage{mathtools}
\usepackage{xcolor}

\newcommand\be{\begin{equation}}
\newcommand\ee{\end{equation}}

\begin{document}

\title 
{\Large \bf Freeze-in Production of Dark Matter Prior to Early Matter Domination} 

\author{Rouzbeh Allahverdi$^{1}$}
\author{Jacek K. Osi{\'n}ski$^{1}$}

\affiliation{$^{1}$~Department of Physics and Astronomy, University of New Mexico, Albuquerque, NM 87131, USA}

\begin{abstract}
Freeze-out or freeze-in during a period of early matter domination can yield the correct dark matter abundance for small values of the velocity-averaged annihilation cross section, $\langle \sigma_{\rm ann} v \rangle_{\rm f} < 3 \times 10^{-26}$ cm$^3$ s$^{-1}$. However, in a generic non-standard thermal history, such a period is typically preceded by other phases. Here, we study production of dark matter in a simple post-inflationary history where a radiation-dominated phase after reheating is followed by an epoch of early matter domination. Focusing on the freeze-in regime, we show that dark matter production prior to early matter domination can dominate the relic abundance in large parts of the parameter space, including weak scale dark matter masses, and the allowed regions are highly dependent on the entire post-inflationary history. Moreover, for a very broad range of $\langle \sigma_{\rm ann} v \rangle_{\rm f}$ spanning over several decades, dark matter particles can start in chemical equilibrium early on and decouple during early matter domination, thereby rendering the relic abundance essentially independent of $\langle \sigma_{\rm ann} v \rangle_{\rm f}$. We briefly discuss connections to different observables as a possible means to test the elusive freeze-in scenario in this case. 
     
\end{abstract}
\maketitle

\section{Introduction}

Despite various lines of evidence for the existence of dark matter (DM)~\cite{BHS}, its identity remains a major problem at the interface of cosmology and particle physics. Weakly interacting massive particles (WIMPs) have long been the focus of direct, indirect, and collider searches for DM. Thermal freeze-out in a radiation-dominated (RD) universe can yield the correct DM abundance if the annihilation rate takes the nominal value $\langle \sigma_{\rm ann} v \rangle_{\rm f} = 3 \times 10^{-26}$ cm$^3$ s$^{-1}$, which is the ballpark value for weak scale masses and interactions (hence called the ``WIMP miracle"). However, this scenario has come under increasing scrutiny by recent experiments, most notably the Fermi-LAT results from observations of dwarf spheroidal galaxies~\cite{fermi1} and newly discovered Milky Way satellites~\cite{fermi2}. A recent analysis~\cite{Beacom} has ruled out thermal DM with a mass below 20 GeV in a model-independent way (unless there is $P$-wave annihilation or co-annihilation), while masses up to 100 GeV can be excluded for specific annihilation channels.

The situation can change if the universe is not in a RD phase at the time of freeze-out~\cite{KT}. An important example is an epoch of early matter domination (EMD), which is a generic feature of early universe models arising from string theory constructions (for a review, see~\cite{KSW}). In this context, an EMD era is driven by long-lived scalar fields (for example, string moduli) that are displaced from the minimum of their potential during inflation, come to dominate the energy density of the post-inflationary universe, and eventually decay to form a RD universe prior to big bang nucleosynthesis (BBN). Both thermal freeze-out and thermal freeze-in during EMD can produce the correct DM relic abundance for small annihilation rates $\langle \sigma_{\rm ann} v \rangle_{\rm f} <3 \times 10^{-26}$ cm$^3$ s$^{-1}$~\cite{GKR,Erickcek}.   

In general, however, an EMD epoch is only one of the stages in the post-inflationary universe. Unless it is driven by oscillations of the inflaton itself, it is typically preceded by a RD phase or a period with a more general equation of state. Thermal freeze-out or freeze-in at these early stages can also contribute to the DM relic abundance (for example, see~\cite{Berlin,Francesco,Erickcek2,Visinelli,Unwin}) and thereby affect the allowed parameter space. Since freeze-out occurs at temperatures below the DM mass $m_\chi$, pre-EMD production will only be relevant for DM masses that are larger than the temperature at the onset of the pre-EMD epoch. On the other hand, in the case of freeze-in, the main contribution to the DM abundance can arise when DM particles are (ultra) relativistic. Thus, pre-EMD production can have a significant affect on the freeze-in side of the parameter space for values of $m_\chi$ at the weak scale or below. In such cases, an exact calculation of the DM relic abundance requires knowledge of the earlier stages of the non-standard thermal history.    

In this work, we perform a detailed study of freeze-in within a non-standard thermal history that involves a RD phase after inflationary reheating followed by a period of EMD.
Such an EMD phase is characterized by two distinct periods: a transition during which the initial radiation energy density redshifts away, which we call the memory phase, and the usual EMD behavior once entropy production becomes significant. We calculate the contributions to the DM relic abundance from production during the prior RD phase as well as the memory phase for the case that $\langle \sigma_{\rm ann} v \rangle_{\rm f}$ is constant over the temperature range of interest. This early contribution to DM production depends on the temperature of the universe at the completion of inflationary reheating $T_{\rm MAX}$, which is the highest temperature of the ensuing RD phase in the post-inflationary universe, and at the onset of the EMD epoch $T_{\rm O}$, in addition to that at the end of the EMD epoch $T_{\rm R}$. We show that pre-EMD production can totally dominate the DM relic abundance in large parts of the $m_\chi - \langle \sigma_{\rm ann} v \rangle_{\rm f}$ plane, and the allowed parameter space    
is highly sensitive to $T_{\rm O}$ and $T_{\rm R}$. A particularly notable observation is that the relic abundance is virtually independent of $\langle \sigma_{\rm ann} v \rangle_{\rm f}$ for a very broad range of $\langle \sigma_{\rm ann} v \rangle_{\rm f}$, spanning over many decades, where DM particles start in chemical equilibrium in the pre-EMD epoch and decouple later on. In this case, measurement of $m_\chi$ at collider experiments, in combination with other cosmological implications of an EMD epoch, may be used as a potential probe of the elusive freeze-in scenario.           

The rest of this paper is organized as follows. In Section II, we discuss a simple non-standard thermal history and calculate the pre-EMD contributions to DM production via freeze-in. In Section III, we present our main results including the allowed regions in the $m_\chi-\langle \sigma_{\rm ann} v \rangle_{\rm f}$ plane and their sensitivity to the history prior to EMD. We elaborate on the correlation between $m_\chi$ and $T_{\rm O}/T_{\rm R}$ in large parts of the parameter space in Section IV, and discuss connections to observables as well as implications for a modulus-driven EMD. We conclude the paper with some discussions in Section V. Some of the details of our calculations are included in the Appendix.

\section{Dark Matter Abundance in the Freeze-in Scenario}

As mentioned, a period of EMD naturally arises in a well-motivated class of early universe models. However, it is typically only one of the stages in the post-inflationary history. We consider a simple scenario that starts with a RD phase at the end of inflationary reheating (for reviews, see~\cite{ABCM,Aminreview}), followed by an EMD epoch driven by oscillations of a long-lived scalar field, or non-relativistic quanta produced in the post-inflationary universe~\cite{Ng,Hooper,Scott2,Cirelli}. A standard RD universe is established at the end of EMD and before BBN. 

Here, we are mainly interested in the evolution of temperature and the freeze-in production of DM without delving into the details of inflationary reheating, the specific particle physics origin of EMD, or the explicit models for DM freeze-in (as done, for example, in~\cite{Bernal}). This can be done by introducing three parameters: the largest temperature in the RD phase after inflationary reheating $T_{\rm MAX}$, the temperature at the onset of EMD $T_{\rm O}$, and the highest temperature in the subsequent RD phase $T_{\rm R}$. The corresponding Hubble expansion rates are denoted by $H_{\rm MAX},~H_{\rm O}$, and $H_{\rm R}$ respectively. 

In calculating the DM relic abundance, without loss of generality, we assume that the DM particle (denoted by $\chi$) represents one degree of freedom. Also, we consider the case where $\langle \sigma_{\rm ann} v \rangle_{\rm f}$ is constant over the temperature range of interest (namely, $T_{\rm R} \lesssim T \lesssim T_{\rm MAX}$)\footnote{Production of DM particles from processes during inflationary reheating that lead to thermalization can be significant (for studies in the case of perturbative reheating, see~\cite{AD1,HMY1,Amin,HMY2}). However, it depends on the details of reheating. This model-dependent contribution, which we do not consider here, can only enhance pre-EMD production of DM and make our conclusions stronger.}. This is the case if DM interacts with particles in the thermal bath via dimension-5 operators without derivative couplings that involve two fermions and two scalar fields. Then, as long as $T$ is below the mass of the mediator that induces these operators, $M$, we have $\langle \sigma_{\rm ann} v \rangle_{\rm f} \propto M^{-2}$. In the Appendix, we give specific examples of such a case. However, in more general situations, $\langle \sigma_{\rm ann} v \rangle_{\rm f}$ can have a temperature dependence\footnote{This is the reason we keep the subscript ${\rm f}$ in $\langle \sigma_{\rm ann} v \rangle_{\rm f}$ throughout the paper. In general, $\langle \sigma_{\rm ann} v \rangle_{\rm f}$ should be interpreted as the value of the annihilation rate at the temperature at which the bulk of the DM relic abundance freezes in.}. In Section V, we will briefly comment on how the $T$ dependence of $\langle \sigma_{\rm ann} v \rangle_{\rm f}$, as well as more general thermal histories involving multiple epochs of EMD separated by RD phases, affect our results. The details of our calculations can be found in the Appendix.

\subsection{The Standard Lore}

Let us briefly recap the well-known case of freeze-in during the late phase of EMD, where decay of the matter component determines the temperature evolution. Deep inside the EMD era, $H_{\rm R} \ll H \ll H_{\rm O}$, there is a subdominant radiation component due to decay of the species that drive EMD. Assuming that the decay products thermalize promptly\footnote{The time scale of thermalization has been estimated, for example, in~\cite{HM}.}, a thermal bath forms with the instantaneous temperature~\cite{GKR,Erickcek}:
\vskip -5mm
\be \label{tinstEMD}
T \approx \left({6 g^{1/2}_{*,{\rm R}} \over 5 g_*}\right)^{1/4} \left({30 \over \pi^2}\right)^{1/8} \left(H T^2_{\rm R} M_{\rm P}\right)^{1/4} , 
\ee
where $g_*$ is the number of relativistic degrees of freedom at temperature $T$, and $M_{\rm P}$ is the reduced Planck mass\footnote{Throughout this paper, the additional subscripts ${\rm R}$, ${\rm MAX}$, and ${\rm O}$ in $g_*$ denote its value at $T_{\rm R}$, $T_{\rm MAX}$, and $T_{\rm O}$ respectively.}. 

In the freeze-in scenario, $\langle \sigma_{\rm ann} v\rangle_{\rm f}$ is so small that DM particles do not reach thermal equilibrium during EMD, and are produced from annihilations of the SM particles. The main contribution to the DM abundance occurs when $T \sim m_\chi/4$~\cite{GKR,Erickcek}. The reason being that particles produced at higher temperatures are quickly diluted by the Hubble expansion, while production at lower temperatures is Boltzmann suppressed. The relic abundance from freeze-in during EMD is given by~\cite{GKR,Erickcek}:
\vskip -5mm
\be \label{EMDdens}
\left(\Omega_{\chi} h^2 \right)_{\rm EMD} \simeq 0.06 ~ {g^{3/2}_{*,{\rm R}} \over g^3_{*,m_\chi/4}} ~  \left({\langle \sigma_{\rm ann} v \rangle_{\rm f} \over 10^{-36} ~ {\rm cm}^3 ~ {\rm s}^{-1}}\right) ~ \left({150 T_{\rm R} \over m_\chi}\right)^5  \left({T_{\rm R} \over 5 ~ {\rm GeV}}\right)^2  ,
\ee
where $g_{*,m_\chi/4}$ is the number of relativistic degrees of freedom at $T = m_\chi/4$.

For a given DM mass, the maximum value of $\langle \sigma_{\rm ann} v \rangle_{\rm f}$ in the freeze-in regime can be approximately found by setting the DM number density $n_\chi$ equal to its thermal equilibrium value $n_{\chi,{\rm eq}}$ at $T \sim m_\chi /4$. For larger values of $\langle \sigma_{\rm ann} v \rangle_{\rm f}$, DM particles reach equilibrium with the thermal bath and hence production transitions to the freeze-out regime. For $T_{\rm R} \sim$ (1-10) GeV, the maximum value of $\langle \sigma_{\rm ann} v \rangle_{\rm f}$ in the freeze-in regime lies within the $10^{-33}-10^{-32}$ cm$^3$ s$^{-1}$ range~\cite{Erickcek}.

\subsection{Production Prior to Early Matter Domination}

In general, an EMD period can be more complicated than that in the previous Subsection. Particularly, if the abundance of radiation during EMD is ever significantly larger than the contribution from decay, then the relation between $T$ and $H$ will deviate, for a time, from that in Eq.~(\ref{tinstEMD}). The temperature evolution throughout the full EMD era may thus have multiple phases. A large abundance of radiation can arise from the presence of additional decaying components (see, for example,~\cite{AO}), or, more simply, from a RD phase that precedes EMD. In the post-inflationary history we are considering, a period of RD is present before EMD resulting in a large abundance of radiation at the onset of EMD\footnote{The effect of substantial initial radiation at early stages of EMD is explored in~\cite{Drees}.}. This radiation then redshifts away until the decay contribution becomes dominant, recovering Eq.~(\ref{tinstEMD}). The temperature evolution during the full EMD period therefore has two phases: an initial phase during which the memory of the prior radiation is being erased, followed by the usual decay-driven phase. 

We now consider the evolution of temperature in the two periods prior to the entropy-producing phase of EMD. 
\vskip 2mm 
\noindent
{\bf RD phase prior to EMD:} The universe is in a RD period for $H_{\rm O} \lesssim H \lesssim H_{\rm MAX}$, during which radiation simply redshifts and temperature is inversely proportional to the scale factor: $T \propto a^{-1}$. This results in the standard relation between $T$ and $H$: 
\vskip -7mm
\be \label{tinstrad}
T = \left({90 \over \pi^2 g_*}\right)^{1/4} \left(H M_{\rm P}\right)^{1/2}. 
\ee
%
\noindent
{\bf Memory phase of EMD:} The radiation and matter components have comparable energy densities at $H = H_{\rm O}$, which signals the beginning of the EMD era\footnote{This is, of course, a continuous transition, but the time of comparable energy densities is nevertheless a good approximation to the beginning of EMD.}. Although $H \propto a^{-3/2}$ for $H_{\rm R} \lesssim H \lesssim H_{\rm O}$, the existing radiation dominates over the contribution from the decaying matter component(s) driving EMD for some time, and continues to redshift. As a result, $T \propto a^{-1}$ for $H_{\rm tran} \lesssim H \lesssim H_{\rm O}$, where:
\be \label{htran1}
H_{\rm tran} \simeq \left({\pi^2 g^{3/5}_{*,{\rm R}} ~ g^{2/5}_{*,{\rm O}} \over 90}\right)^{1/2} {T^{6/5}_{\rm R} T^{4/5}_{\rm O} \over M_{\rm P}},
\ee
and the relation between $T$ and $H$ is\footnote{Eq.~(\ref{tinsttran}) is easily obtained by evaluating Eq.~(\ref{tinstrad}) at $T = T_{\rm O}$ and using the appropriate redshift relation for memory EMD, while Eq.~\ref{htran1} makes use of the redshift relations all the way to $T = T_{\rm R}$. One can also obtain Eqs.~(\ref{htran1},\ref{tinsttran}) from Eq.~(9) of~\cite{AO} with $f = 1$ and $\alpha = (g_{*,{\rm O}} T^4_{\rm O}/g_{*,{\rm R}} T^4_{\rm R})^{1/2}$ to match the case under consideration.}:
\vskip -6mm
\be \label{tinsttran}
T \approx \left({90 \over \pi^2 g^{3/4}_* g^{1/4}_{*,{\rm O}}}\right)^{1/3} \left({H^2 M^2_{\rm P} \over T_{\rm O}}\right)^{1/3}.
\ee
Once $H \ll H_{\rm tran}$, the memory of the initial radiation is completely erased and the $T$ dependence on $H$ transitions to that of Eq.~(\ref{tinstEMD}) until the end of EMD. 

\vskip 2mm
\begin{figure}[ht!]
  \centering
  \includegraphics[width=0.5\textwidth]{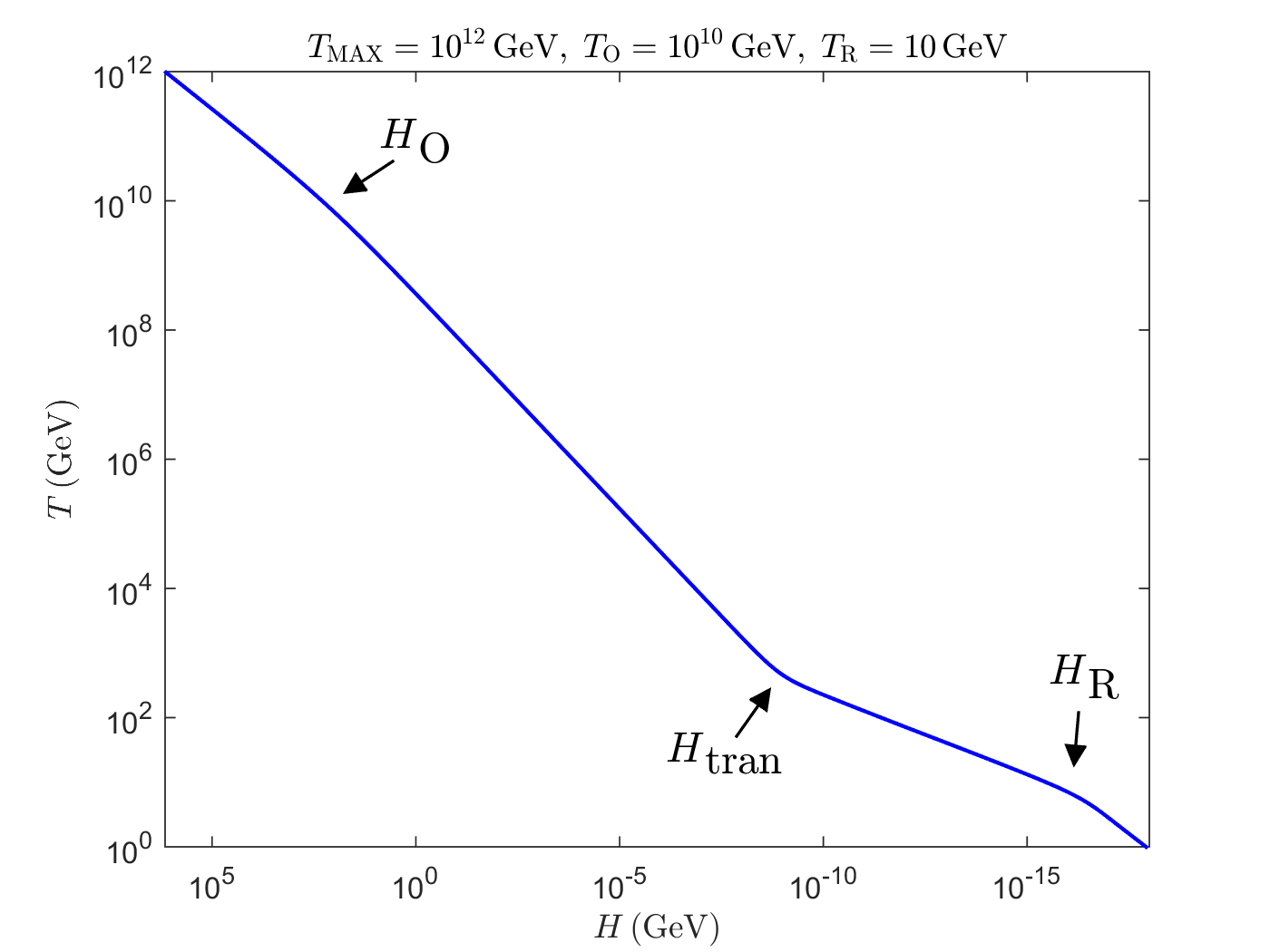}
  \captionsetup{justification = raggedright}
  \caption{Temperature of the universe as a function of the Hubble expansion rate showing the post-inflationary history we are considering. The curve begins at the top left corner, when \(H \approx H_{\rm MAX}\), with an early period of RD. The EMD period lasts from \(H \approx H_{\rm O}\) to \(H \approx H_{\rm R}\), with $H_{\rm tran} \lesssim H \lesssim H_{\rm O}$ corresponding to the memory phase, and $H_{\rm R} \lesssim H \lesssim H_{\rm tran}$ to the entropy-producing phase. The universe returns to a RD phase after the end of EMD, where \(H < H_{\rm R}\). }
  \label{fig:1}
\end{figure}
We show the evolution of $T$ in terms of $H$ in Fig.~\ref{fig:1}, derived from numerical calculations, through the entire non-standard thermal history considered here for $T_{\rm MAX} = 10^{12}$ GeV, $T_{\rm O} = 10^{10}$ GeV, and $T_{\rm R} = 10$ GeV. The evolution during the different stages is in very good agreement with the relations given in Eqs.~(\ref{tinstEMD},\ref{tinstrad},\ref{tinsttran}). 

As shown in the Appendix, the relation between $T$ and $H$ in these two phases implies that the main contribution to DM production occurs at the highest temperature in each phase. Hence, with $T_{\rm MAX} \gg T_{\rm O}$, the DM relic abundance is set in the pre-EMD epoch. 
This pre-EMD component of the relic abundance has two distinct regimes based on the value of $\langle \sigma_{\rm ann} v \rangle_{\rm f}$. If it is very small, DM particles will not be able to establish chemical equilibrium throughout the post-inflationary history.
Production in this decoupling regime will dominantly occur at $T \simeq T_{\rm MAX}$. If $\langle \sigma_{\rm ann} v \rangle_{\rm f}$ is large enough, then 
DM particles will reach chemical equilibrium in the pre-EMD phase but can decouple at $H \gtrsim H_{\rm tran}$, hence the early-equilibrium regime. This regime simply refers to the situation where chemical decoupling of particles occurs while they are relativistic. 
The condition on $\langle \sigma_{\rm ann} v \rangle_{\rm f}$ to be in either of these regimes is given in Eqs~(\ref{ficond},\ref{deccond}) of the Appendix:
\vskip -6mm
\begin{eqnarray} \label{preEMDcond}
&&\langle \sigma_{\rm ann} v \rangle_{\rm f} \ll {\pi^3 g^{1/2}_{*,{\rm MAX}} \over \sqrt{90} \zeta(3) M_{\rm P} T_{\rm MAX}} ~~~~~~~~~~~~~~~ {\rm {\bf (decoupling ~ regime)}} \, , \nonumber \\ 
&&\langle \sigma_{\rm ann} v \rangle_{\rm f} \gtrsim {\pi^3 g^{1/2}_{*,{\rm MAX}} \over \sqrt{90} \zeta(3) M_{\rm P} T_{\rm MAX}} ~~~~~~~~~~~~~~~ {\rm {\bf (early\text{-}equilibrium ~ regime)}}\, . 
\end{eqnarray}
The pre-EMD contribution to the DM abundance follows from Eqs.~(\ref{fifinal},\ref{decfinal}):
\begin{equation} \label{preEMDdens}
\left(\Omega_\chi h^2 \right)_{\rm pre\text{-}EMD} \simeq
\begin{dcases}
{0.028 \over g^{5/4}_{*,{\rm MAX}} g^{1/4}_{*,{\rm O}}} \langle \sigma_{\rm ann} v \rangle_{\rm f} M_{\rm P}  \left({m_\chi \over 1 ~ {\rm GeV}} \right) \left({10^9 T_{\rm MAX} T_{\rm R} \over T_{\rm O}}\right) & ~~~~~ {\rm {\bf (decoupling ~ regime)}}
\, , \\ 
{0.076 \over g_{*,{\rm dec}}} \left({m_\chi \over 1 ~ {\rm GeV}}\right) \left({10^9 T_{\rm R} \over T_{\rm O}}\right) & ~~~~~ {\rm {\bf (early\text{-}equilibrium ~ regime)}}
\, . 
    \end{dcases}
\end{equation}
%
%

A comment is in order at this point. The early-equilibrium regime simply refers to the situation where chemical decoupling of DM particles occurs while they are relativistic. A well-known example of this is the decoupling of neutrinos from the thermal bath at $T \sim {\cal O}({\rm MeV})$. However, this happens around BBN when the universe is in a RD phase. This case can be obtained by taking $T_{\rm R} = T_{\rm O}$, corresponding to the absence of an EMD epoch, in the scenario discussed here. Taking $T_{\rm R} = T_{\rm O}$, and using $g_{*,{\rm dec}} = 10.75$, the second expression in Eq.~(\ref{preEMDdens}) reproduces the correct DM abundance for $m_\chi \simeq 17$ eV. This is in agreement with $\sum{m_\nu} \simeq 11$ eV, obtained from the well-known relation for the sum of the neutrino masses $\Omega_\nu h^2 = \sum{m_\nu}/93 ~ {\rm eV}$, after dividing by $3/2$ to account for the two fermionic degrees of freedom for a neutrino instead of a single one for $\chi$ which we have assumed here.

Let us now underline some important differences between the pre-EMD and EMD components of the relic abundance given in Eqs.~(\ref{preEMDdens}) and~(\ref{EMDdens}) respectively:  
%
\vskip 2mm
\begin{itemize}
\item The pre-EMD component depends on $T_{\rm O}$ and $T_{\rm MAX}$ in addition to $T_{\rm R}$. Freeze-in during pre-EMD is most efficient at the highest temperature in that era, which explains the appearance of $T_{\rm MAX}$ in the first expression in Eq.~(\ref{preEMDdens}). Since DM particles start in chemical equilibrium in the early-equilibrium regime, the second expression in~(\ref{preEMDdens}) is independent from $T_{\rm MAX}$. The factor $T_{\rm R}/T_{\rm O}$ appears in both cases indicating dilution by entropy generation during the EMD epoch. 
    
\item The pre-EMD component is proportional to $m_\chi$. This can be understood by noting that the bulk of DM production in the pre-EMD phase occurs when $T \gg m_\chi$. Therefore, the resulting DM number density, see Eqs.~(\ref{fifinal},\ref{decfinal}), is independent of $m_\chi$. 
    
\item The pre-EMD component has no explicit dependence on $\langle \sigma_{\rm ann} v \rangle_{\rm f}$ in the early-equilibrium regime, the second expression in Eq.~(\ref{preEMDdens}), because DM particles start in chemical equilibrium. The only role of $\langle \sigma_{\rm ann} v \rangle_{\rm f}$  in this case is to determine the decoupling temperature $T_{\rm dec}$, which results in an implicit dependence through $g_{*,{\rm dec}}$. 
    
\item The pre-EMD component has a much milder dependence on $T_{\rm R}$ and $m_\chi$ than the EMD component. As a result, moderate changes in these parameters can render the latter totally negligible, significantly affecting the allowed parameter space. This will become clear when we present our results in the next Section.    
\end{itemize}

\section{Results}

In this Section, we present our results. We numerically solve the set of Boltzmann equations governing the evolution of radiation, ${\rm r}$, the decaying component driving EMD, $\phi$, and DM particles, $\chi$: 
\begin{eqnarray}\label{boltzmann}
&& {\dot \rho_{\rm r}} + 4H\rho_{\rm r} = \Gamma_\phi\rho_\phi - \langle E_{\chi}\rangle\langle\sigma_{\rm ann} v \rangle_{\rm f} \left(n^2_{\chi,{\rm eq}} - n^2_\chi\right) \nonumber \\[2mm]
&& {\dot \rho_\phi} + 3H\rho_\phi = - \Gamma_\phi\rho_\phi \\[2mm] 
&& {\dot n}_\chi + 3 H n_\chi =  \langle \sigma_{\rm ann} v \rangle_{\rm f}  \left(n^2_{\chi,{\rm eq}} - n^2_\chi \right) \nonumber 
\end{eqnarray}
where $\Gamma_\phi$ is the decay rate of $\phi$, $\langle E_{\chi}\rangle \approx \left(m_\chi^2 + 9T^2\right)^{1/2}$ is the average energy per DM particle\footnote{The contribution of this term to the radiation energy density is typically very small, even when $n_{\chi,{\rm eq}} \gg n_\chi$, and hence the exact form of $\langle E_{\chi}\rangle$ is not important for the overall evolution.}, and $n_{\chi,{\rm eq}}$ denotes the thermal equilibrium value of the DM number density. The energy density in $\phi$ is tracked until it is sufficiently small to be unimportant for the subsequent evolution, and is then dropped to facilitate faster numerical calculation. We have taken the detailed temperature dependence of the \(g_*\) factor into account down to well below $T_{\rm R}$. In order to calculate the DM relic abundance, we normalize the DM number density with the entropy density long after the end of the EMD epoch.

We investigate the parameter space in the $m_\chi-\langle \sigma_{\rm ann} v \rangle_{\rm f}$ plane that yields the correct DM abundance for various values of $T_{\rm R}$, $T_{\rm O}$, and $T_{\rm MAX}$. A given $m_\chi$ can result in an overabundance or underabundance of DM for different values of $\langle \sigma_{\rm ann} v \rangle_{\rm f}$, and the values that reproduce the correct DM abundance depend on the thermal history. We therefore focus on the values that yield $\Omega_\chi h^2 = 0.12$ in order to study this dependence. Additionally, in our results below, we use the contribution to the relic abundance from the entropy-producing phase of EMD as a baseline for comparison. 

\begin{figure}[ht!]
  \centering
  \includegraphics[width=0.5\textwidth]{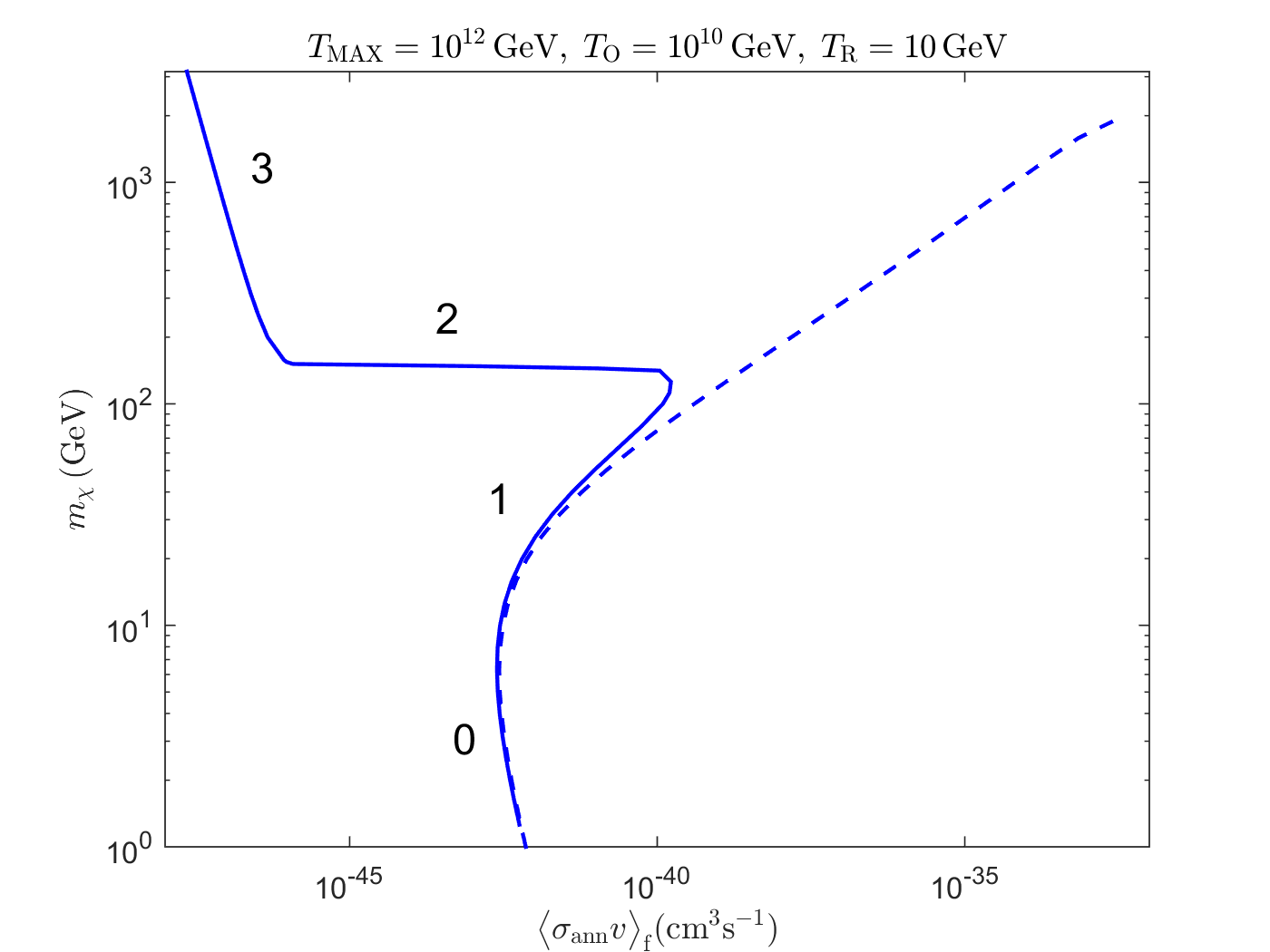}
  \captionsetup{justification = raggedright}
  \caption{Values of $m_\chi$ and $\langle \sigma_{\rm ann} v \rangle_{\rm f}$ that yield the DM relic abundance of $\Omega_\chi h^2 \approx 0.12$, obtained by numerical solution of the Boltzmann equations \eqref{boltzmann}. The solid curve corresponds to production during the entire thermal history shown in Fig.~\ref{fig:1}, while the dashed curve depicts the freeze-in side of the corresponding baseline curve from production in the entropy-generating phase of EMD as well as the RD phase that follows. For a fixed $m_\chi$, DM is overproduced for larger values of $\langle \sigma_{\rm ann} v \rangle_{\rm f}$ than those given by the curves, while smaller values result in underproduction. In region 0, the dominant DM contribution arises from freeze-in during the RD phase after the end of EMD; in region 1, the DM relic abundance is dominated by the late EMD contribution; while the early-equilibrium and decoupling regimes of the pre-EMD contribution dominate in regions 2 and 3, respectively. The transition to freeze-out of the baseline curve is seen to begin at the top-right corner. }
  \label{fig:2}
\end{figure}

In Fig.~\ref{fig:2}, we show the curve in the $m_\chi-\langle \sigma_{\rm ann} v \rangle_{\rm f}$ plane that represents points for which 
the observed DM abundance is obtained for $T_{\rm MAX} = 10^{12}\,{\rm GeV}$, $T_{\rm O} = 10^{10}\,{\rm GeV}$, and $T_{\rm R} = 10\,{\rm GeV}$. We include a short segment (region 0) where the dominant DM contribution comes from production after the end of EMD for reference. The rest of the curve consists of three distinct regions, 1, 2, and 3, which correspond to DM production during and before the EMD phase as follows:
\vskip 2mm
\noindent
\begin{enumerate}
\item[${\bf (1)}$] Region 1 starts at DM masses $m_\chi \approx T_{\rm R}$ and initially follows the baseline curve, but moves above it as $m_\chi$ and $\langle \sigma_{\rm ann} v \rangle_{\rm f}$ increase. This behavior can be understood from the analytical approximations in Eqs.~(\ref{EMDdens},\ref{preEMDdens}). The EMD component dominates at small masses due to its scaling $\propto m^{-5}_\chi$. As $m_\chi$ and $\langle \sigma_{\rm ann} v \rangle_{\rm f}$ both increase along the baseline curve, the pre-EMD component~(\ref{preEMDdens}) becomes more relevant. Obtaining the correct DM abundance then requires a larger $m_\chi$ than the EMD component alone, and hence the curve goes above the baseline.    

\item[${\bf (2)}$] Region 2 starts at the turning point (the abrupt departure from the baseline curve) and extends to very small values of $\langle \sigma_{\rm ann} v \rangle_{\rm f}$. It is essentially horizontal for the following reason. In this region, the pre-EMD component dominates and $\langle \sigma_{\rm ann} v \rangle_{\rm f}$ is large enough that we are in the early-equilibrium regime. The relic abundance is then given by the second expression in Eq.~(\ref{preEMDdens}), which is almost independent from $\langle \sigma_{\rm ann} v \rangle_{\rm f}$. However, region 2 is not exactly horizontal as $\left(\Omega_\chi h^2 \right)_{\rm EMD}$ brings in a very mild $m_\chi$ dependence that is too small to be noticeable in the figure. 

\item[${\bf (3)}$] Region 3 starts at the point where $\langle \sigma_{\rm ann} v \rangle_{\rm f} \!\sim 3 g^{1/2}_{*,{\rm MAX}} \left(T_{\rm MAX} M_{\rm P}\right)^{-1}$, which corresponds to the boundary between the two regimes of Eq.~\eqref{preEMDcond}, and rises toward larger values of $m_\chi$ as $\langle \sigma_{\rm ann} v \rangle_{\rm f}$ decreases. In this region, $\langle \sigma_{\rm ann} v \rangle_{\rm f}$ is so small that production of DM particles in the pre-EMD phase occurs in the decoupling regime. As a result, the DM relic abundance is dominated by the pre-EMD component and follows the first expression in Eq.~(\ref{preEMDdens}). 

\end{enumerate}
%

\begin{figure}[ht!]
  \centering
  \subfloat{\includegraphics[width=0.5\textwidth]{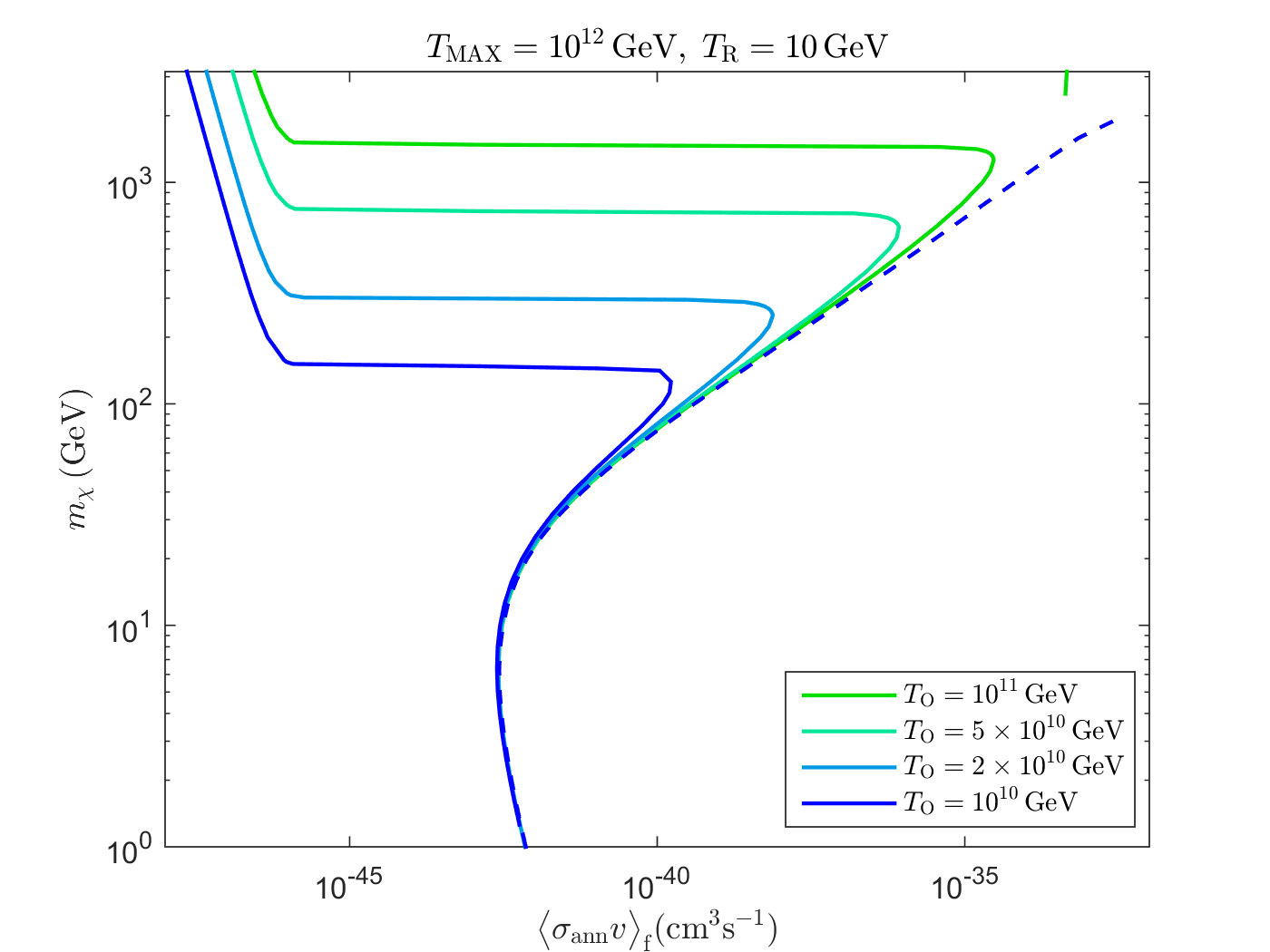}}
  \subfloat{\includegraphics[width=0.5\textwidth]{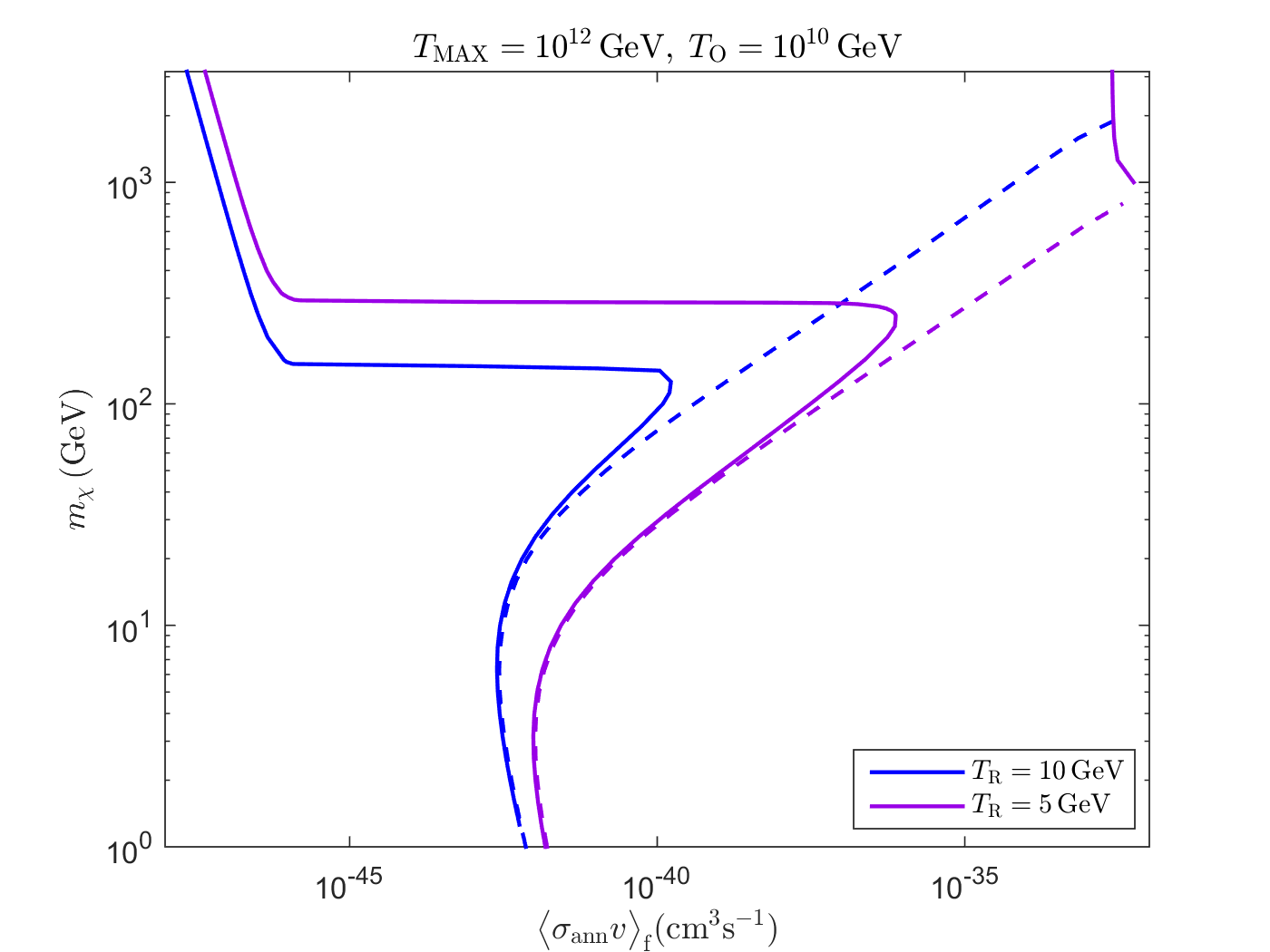}}
  \captionsetup{justification = raggedright}
  \caption{Variation of the curve from Fig.~\ref{fig:2} for different values of \(T_{\rm R}\) and \(T_{\rm O}\). Left: variation of \(T_{\rm O}\) for constant \(T_{\rm MAX}\) and \(T_{\rm R}\). Right: variation of \(T_{\rm R}\) for constant \(T_{\rm MAX}\) and \(T_{\rm O}\). Note the appearance of the freeze-out side, at the top right corner, which merges with the peak of the baseline curve. }
  \label{fig:3}
\end{figure}

\begin{figure}[ht!]
  \centering
  \includegraphics[width=0.5\textwidth]{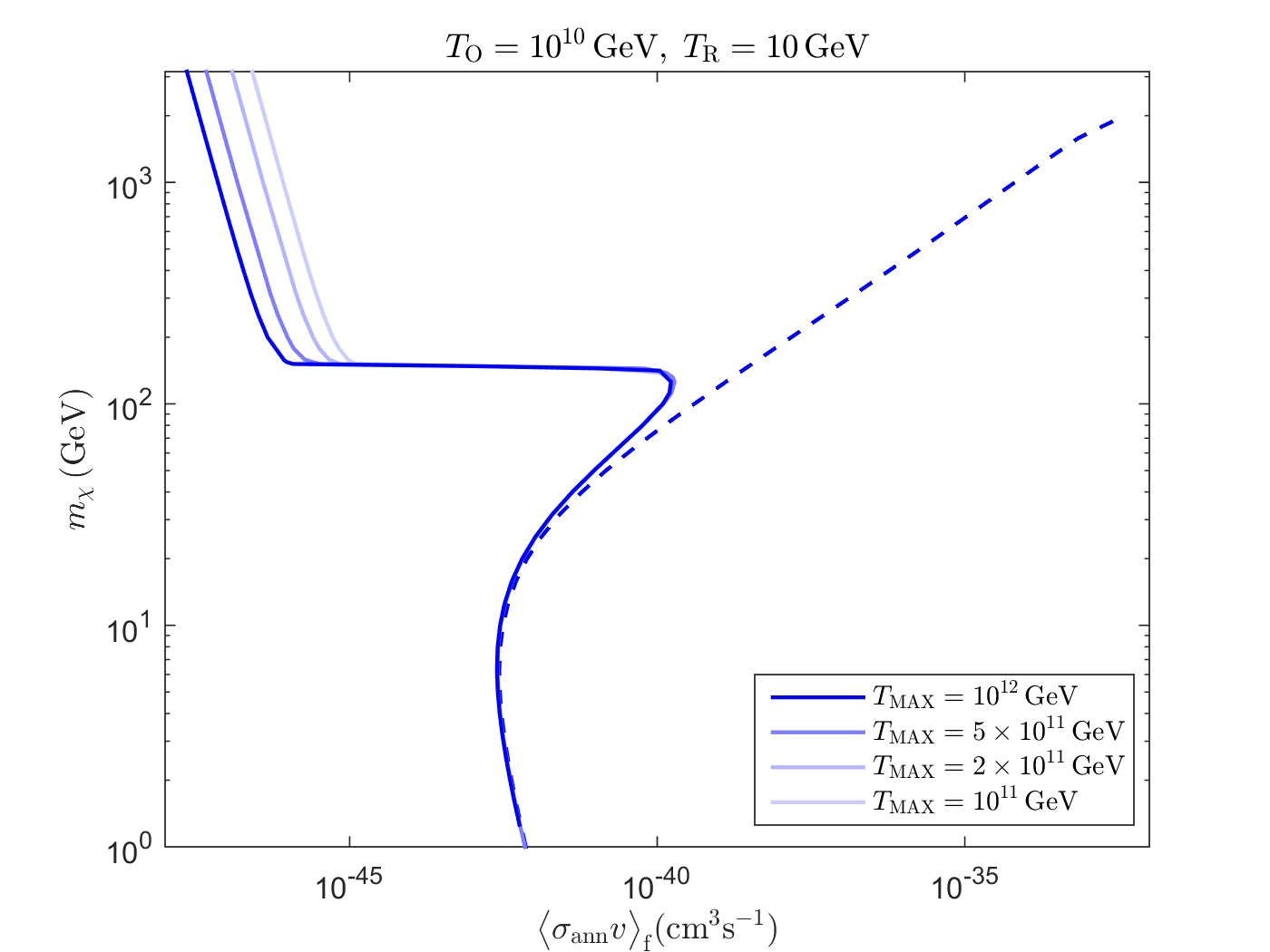}
  \caption{Variation of the curve from Fig.~\ref{fig:2} for different values of \(T_{\rm MAX}\) while holding \(T_{\rm R}\) and \(T_{\rm O}\) constant. }
  \label{fig:4}
\end{figure}

Figs.~\ref{fig:3} and \ref{fig:4} depict the sensitivity of the allowed parameter space on the post-inflationary history. In Fig.~\ref{fig:3}, we show variation of the curve from Fig.~\ref{fig:2} for different values of $T_{\rm O}$ for fixed $T_{\rm R}$ and $T_{\rm MAX}$ (left panel), and for different values of $T_{\rm R}$ when $T_{\rm O}$ and $T_{\rm MAX}$ are kept constant (right panel). Fig.~\ref{fig:4}, shows the change in the curve when $T_{\rm MAX}$ is varied for fixed $T_{\rm R}$ and $T_{\rm O}$. We observe the following main features in the figures:
\vskip 2mm
\noindent
\begin{itemize}
\item In Fig.~3, region 2 moves up with increasing $T_{\rm O}$ and down with increasing $T_{\rm R}$, while there is no change when $T_{\rm MAX}$ varies in Fig.~4. Since pre-EMD production in the early-equilibrium regime dominates in this region, we expect $m_\chi \propto T_{\rm O}/T_{\rm R}$, independent of $T_{\rm MAX}$, which agrees with what we observe in the figures.     
    
\item The turning point is highly dependent on $T_{\rm R}$ and $T_{\rm O}$, as seen in Fig.~3, but does not change with $T_{\rm MAX}$ in Fig.~4. Its position can be estimated by setting the sum of the EMD component~(\ref{EMDdens}) and the second expression in Eq.~(\ref{preEMDdens}) equal to the observed DM abundance and finding the local maximum of $\langle \sigma_{\rm ann}v \rangle_{\rm f}$ in terms of $m_\chi$. As it turns out, $\langle \sigma_{\rm ann} v \rangle_{\rm f} \propto T^5_{\rm O} T^{-12}_{\rm R}$ at the turning point. This is in agreement with the considerable horizontal movement (specially when $T_{\rm R}$ changes) in Fig.~\ref{fig:3}. As expected, the vertical shift follows that of region 2 mentioned above.  
    
\item The left end of region 2, where it meets region 3, moves horizontally with changing $T_{\rm MAX}$ in Fig.~4, and vertically when $T_{\rm O}$ and $T_{\rm R}$ are varied in Fig.~3. This point divides the decoupling and early-equilibrium regimes of the pre-EMD contribution, where we have $\langle \sigma_{\rm ann} v \rangle_{\rm f} \propto T^{-1}_{\rm MAX}$, which explains the horizontal movement. As expected, the dependence on $T_{\rm O}$ and $T_{\rm R}$ follows that of region 2. 
\end{itemize}
%

A very important point is that pre-EMD production opens up vast regions of the parameter space at very small $\langle \sigma_{\rm ann} v \rangle_{\rm f}$ that are not allowed when the EMD component alone is considered. In fact, the freeze-in side of the baseline curve does not extend below a certain value of $\langle \sigma_{\rm ann} v \rangle_{\rm f}$ corresponding to the transition between regions 1 and 0 shown in Fig.~\ref{fig:2}. This is because lowering $\langle \sigma_{\rm ann} v \rangle_{\rm f}$ results in a smaller $m_\chi$ in order to obtain the correct relic abundance. However, once $T_{\rm f} \sim m_\chi/4$ drops below $T_{\rm R}$, the contribution to the relic abundance from RD after EMD becomes important. This sets a lower bound on $\langle \sigma_{\rm ann} v \rangle_{\rm f}$, for a given $T_{\rm R}$, beyond which the baseline curve does not extend. Nevertheless, the pre-EMD contribution can still dominate for such small $m_\chi$ and $\langle \sigma_{\rm ann} v \rangle_{\rm f}$, especially for combinations of the parameters that lower region 2, such as $T_{\rm MAX} = 10^{12}\,{\rm GeV}$, $T_{\rm O} = 10^{8}\,{\rm GeV}$, and $T_{\rm R} = 10\,{\rm GeV}$. 

For the values of $T_{\rm O}$ and $T_{\rm R}$ in Figs.~\ref{fig:2}-\ref{fig:4},
the pre-EMD component leads to a separation of the freeze-in and freeze-out parts of the allowed parameter space by a horizontal gap. Though the freeze-out part is not affected as much as the freeze-in part, and generally lies to the right of the figures, we include a short segment at the top right corner of Fig.~2 for reference. Decreasing $T_{\rm R}$ and/or increasing $T_{\rm O}$ results in a growing overlap between region 1 and the baseline curve, and a significant movement of the turning point to the right, making region 2 larger. However, the turning point cannot go beyond the peak of the baseline curve because larger values of $\langle \sigma_{\rm ann} v \rangle_{\rm f}$ actually give rise to freeze-out. Similarly, the freeze-out side moves toward the left and up with increasing $T_{\rm O}/T_{\rm R}$. 

\begin{figure}[ht!]
  \centering
  \subfloat{\includegraphics[width=0.5\textwidth]{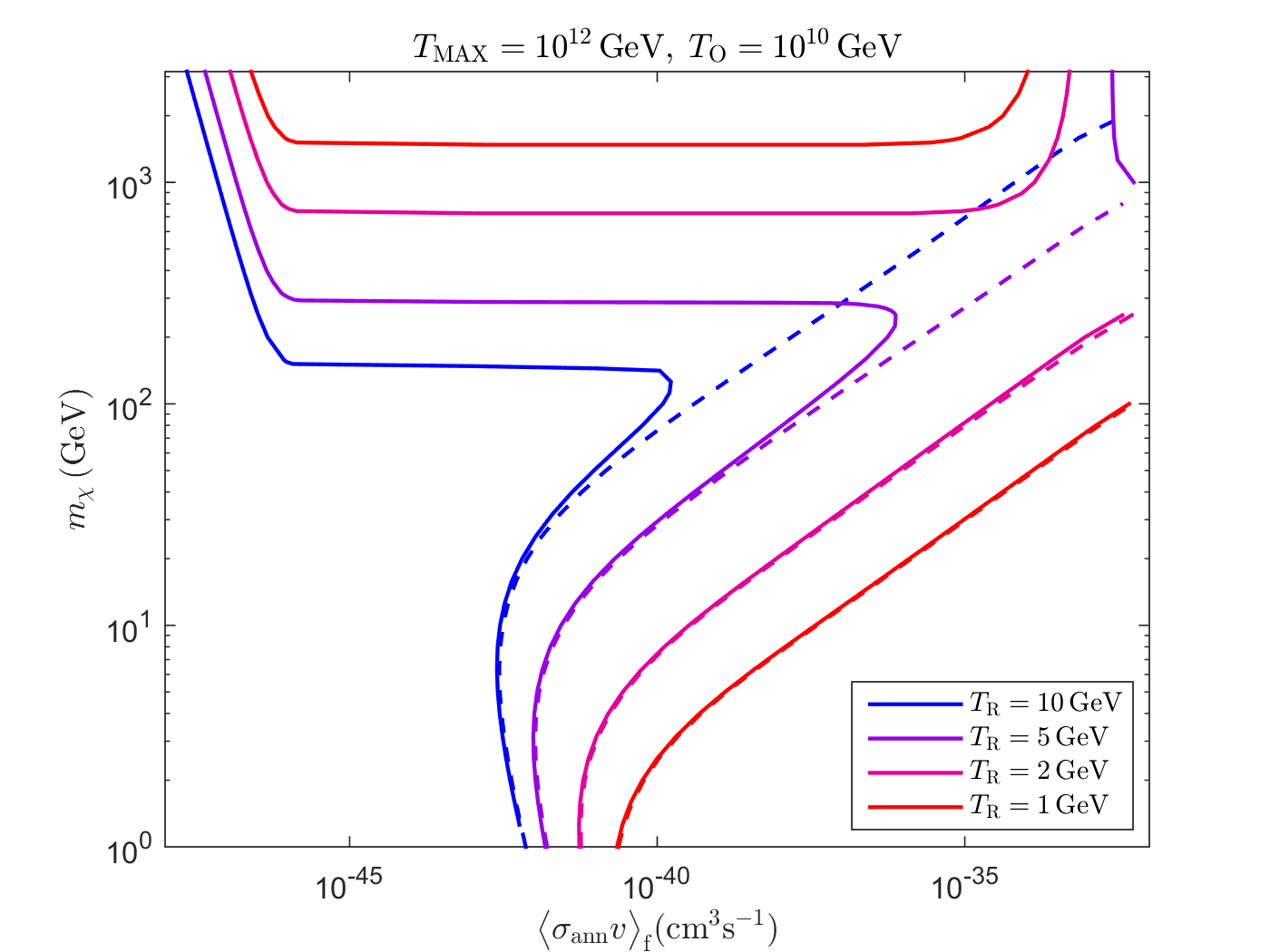}}
  \subfloat{\includegraphics[width=0.5\textwidth]{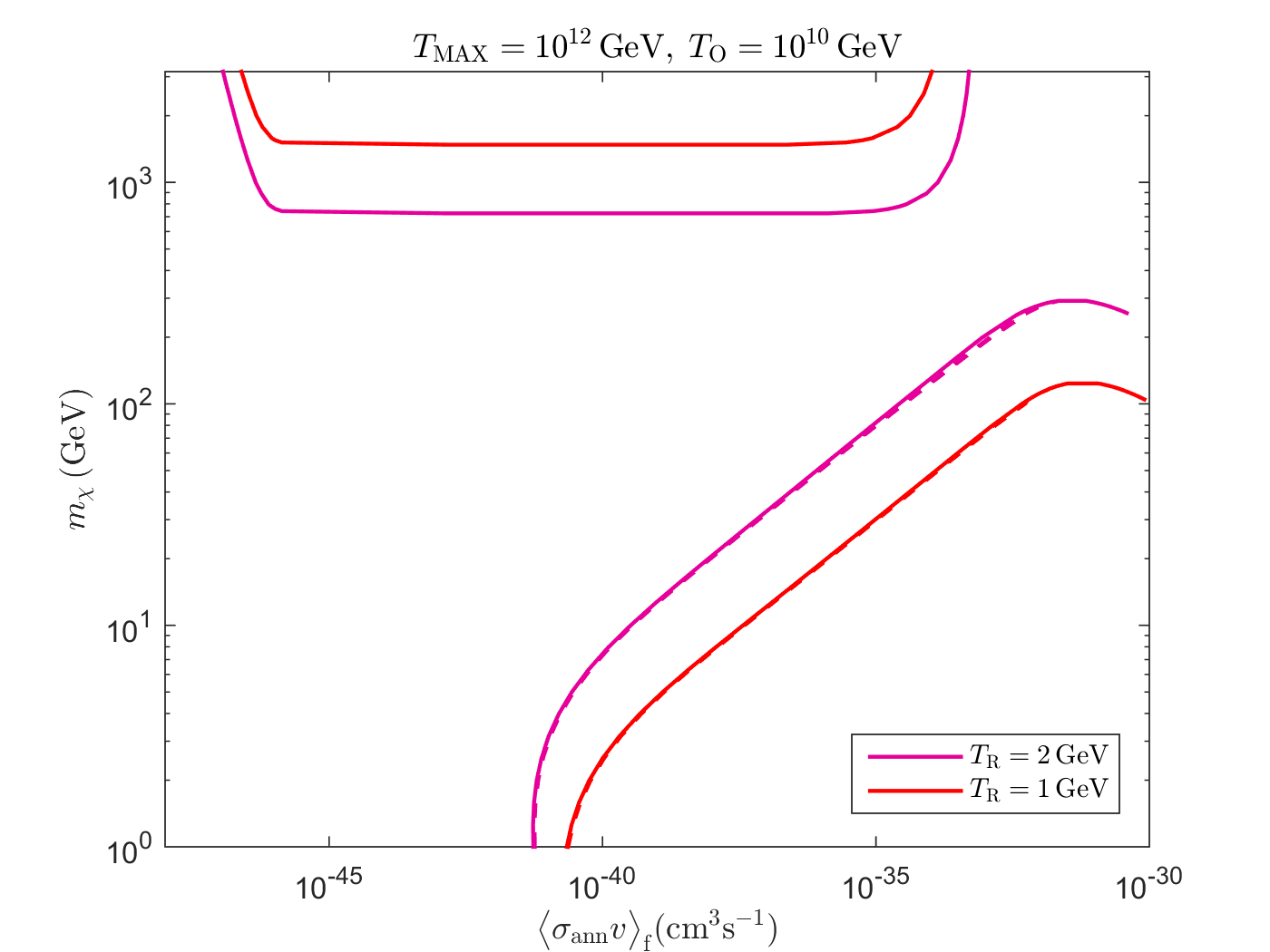}}
  \captionsetup{justification = raggedright}
  \caption{Left: variation of \(T_{\rm R}\), with constant \(T_{\rm MAX}\) and \(T_{\rm O}\), for values that display the transition between the horizontal gap and vertical gap behavior described in the text. For the larger two values of \(T_{\rm R}\), which are taken from the right panel of Fig.~\ref{fig:3}, the solid curves are separated into a freeze-in part on the left and a freeze-out part on the right (which we do not show apart from the top right corner). For the two smaller values, the curves are split into a lower part that follows the baseline curve and an upper part that connects the freeze-in and freeze-out sides. Right: the transition to freeze-out is shown at slightly larger annihilation rates for the two smaller reheat temperatures from the left panel.}
  \label{fig:5}
\end{figure}

At some point, the freeze-in and freeze-out parts join and region 1 coincides with the freeze-in side of the baseline curve, while region 2 splits away. By further increasing $T_{\rm O}/T_{\rm R}$, the allowed parameter space is again divided into two disjointed parts that are now separated by a vertical gap. We clearly see this in Fig.~\ref{fig:5} where the smaller two values of $T_{\rm R}$ have fully split from the baseline curve. For these two, region 1 follows the full baseline curve, including the freeze-out side which is partially shown in the right panel. The upper segments include regions 2 and 3, discussed above, that make a smooth transition to the freeze-out regime rising up on the right side. In summary, at lower $T_{\rm O}/T_{\rm R}$, the effect of pre-EMD production of DM is to disconnect the allowed parameter space into the freeze-in and freeze-out parts, while for higher $T_{\rm O}/T_{\rm R}$, it gives rise to a new curve that smoothly interpolates between the freeze-in and freeze-out regimes but is situated at larger values of $m_\chi$ compared to the baseline curve.

\section{Early-Equilibrium Regime: A Closer Look}

A remarkable feature observed in Figs.~2-5 is that the DM relic abundance is essentially independent from $\langle \sigma_{\rm ann} v \rangle_{\rm f}$ in large parts of the parameter space. This is due to the fact that DM particles start in chemical equilibrium during the pre-EMD phase for a broad range of $\langle \sigma_{\rm ann} v \rangle_{\rm f}$. This range, corresponding to the early-equilibrium regime, spans over many orders of magnitude extending from a minimum value determined by $T_{\rm MAX}$, see Eq.~(\ref{deccond}), to a maximum value that depends on $T_{\rm O}$ and $T_{\rm R}$ and can be as large as that at the peak of the baseline curve. This feature can be considered as the freeze-in analogue to the WIMP miracle in a complementary way: while the relic abundance mainly depends on $\langle \sigma_{\rm ann} v \rangle_{\rm f}$ for the latter, it is mostly dependent on $m_\chi$ in this case. 

The DM abundance in this case is given by the second expression in Eq.~(\ref{preEMDdens}). This is at most equal to the observed relic abundance as, in general, there exist other sources that contribute to the total abundance. Most notably, DM particles may be directly produced in the decay of the field(s) driving the EMD epoch~\cite{KMY,MR,GG,ADS}. As a result, we find the following inequality in order to not overproduce DM:
\be \label{ub}
m_\chi \lesssim 1.6 \; g_{*,{\rm dec}} \left({T_{\rm O} \over 10^9 T_{\rm R}}\right) (1~ {\rm GeV}) .
\ee

\begin{figure}[ht!]
  \centering
  \includegraphics[width=0.5\textwidth]{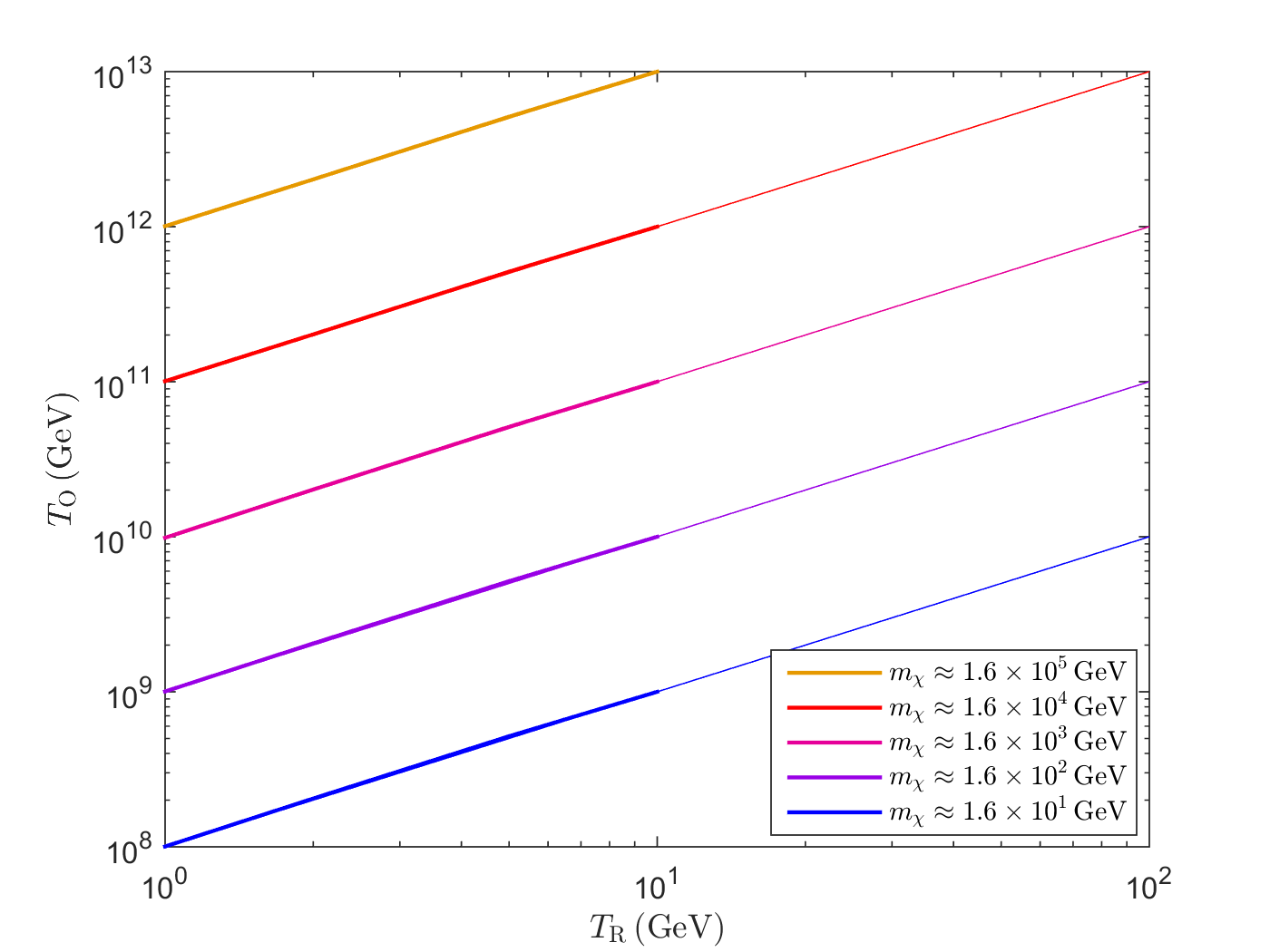}
  \captionsetup{justification = raggedright}
  \caption{Contours of the upper bound on \(m_\chi\) in the $T_{\rm O}-T_{\rm R}$ plane, in order to not overproduce DM in the early-equilibrium regime. The solid lines are obtained numerically from region 2 shown in previous figures, while the thin lines correspond to Eq.~\eqref{ub}.}
  \label{fig:6}
\end{figure}

In Fig.~\ref{fig:6}, we show the contours corresponding to the maximum allowed value of $m_\chi$ in the $T_{\rm O}-T_{\rm R}$ plane. The thick line segments are from full numerical calculations, while the thin lines represent the RH side of~(\ref{ub}). In the latter, we have taken $g_{*,{\rm dec}} = 106.75$, the value for the SM, which is a very good approximation for the range of $T_{\rm R}$ and $T_{\rm O}$ shown in the figure. The numerical and analytical results agree very well for the range of parameters chosen. 

\subsection{Connection to Observables}

If the two sides of~(\ref{ub}) can be connected to experiments, then the inequality can be considered as a consistency relation of the pre-EMD early-equilibrium regime. The LH side is the DM mass, which can in principle be measured at the LHC (or future colliders) through the standard missing energy signal for the range of $m_\chi$ shown in Figs.~\ref{fig:2}-\ref{fig:5}. One may also impose an upper bound on the RH side, which is a direct measure of the duration of the EMD era, in the context of inflationary cosmology. 

In the non-standard history we have considered here, the number of e-foldings of inflation between the time when cosmologically relevant perturbations, corresponding to the pivot scale $k_* = 0.05$ Mpc$^{-1}$, left the horizon and the end of inflation can be written as~\cite{LL, RG}\footnote{The second term on the RH side of Eq.~\eqref{nkt} is basically the second term on the RH side of Eq.~(8) in \cite{LL} written in terms of $r$, where we relate $r$ and the Hubble rate at horizon crossing as in \cite{RG}.}:
\be \label{nkt}
N_{k_*} \approx 57.6 + {1 \over 4} {\rm ln} r - \Delta N_{\rm reh} - \Delta N_{\rm EMD} ,
\ee
where:
\be \label{nkt2}
\Delta N_{\rm reh} \equiv {1 - 3 w_{\rm reh} \over 6 (1 + w_{\rm reh})} ~ {\rm ln} \left({H_{\rm inf} \over H_{\rm MAX}}\right) ~ ~ ~ , ~ ~ ~ \Delta N_{\rm EMD} \equiv {1 \over 6} {\rm ln} \left({H_{\rm O} \over H_{\rm R}}\right).
\ee
Here, $H_{\rm inf}$ is the Hubble rate during inflation, $r$ is the tensor-to-scalar ratio, and $w_{\rm reh}$ represents the equation of state during reheating.  Using the relation between $H$ and $T$ at the onset and the end of the EMD epoch, we can write:
\be \label{nkt3}
\Delta N_{\rm EMD} \simeq {1 \over 3} {\rm ln} \left({T_{\rm O} \over T_{\rm R}}\right) + {1 \over 12} {\rm ln} \left({g_{*,{\rm O}} \over g_{*,{\rm R}}}\right).
\ee

Theoretical arguments and numerical simulations suggest that generally $0 \leq w_{\rm reh} \leq 1/3$~\cite{PFKP,LA}, which implies that $\Delta N_{\rm reh} \geq 0$. Combined with the experimental bound {\bf $r_{\rm max} \simeq 0.064$}~\cite{Planck18}, and the typical range of values for $g_{*,{\rm O}}$ and $g_{*,{\rm R}}$, we find:
\be \label{CMBbound} 
{\rm ln} \left({T_{\rm O} \over T_{\rm R}}\right)  \lesssim 173 - 3 N_{k_*} .
\ee
$N_{k_*}$ is related to the scalar spectral index $n_{\rm s}$ that is constrained by the cosmic microwave background (CMB) experiments. The relation between $N_{k_*}$ and $n_{\rm s}$ is model-dependent as it is determined by the shape of the inflaton potential. However, there are simple relations between $N_{k_*}$ and $n_{\rm s}$~\cite{Roest} in two important universality classes of single field models of inflation that include a large number of models compatible with the latest Planck results~\cite{Planck18}. Therefore, in these cases, one can use the experimental bounds on $n_{\rm s}$ to impose a lower limit on $N_{k_*}$ and, through~(\ref{CMBbound}), an upper limit on $T_{\rm O}/T_{\rm R}$~\cite{ADM}.

In addition, one may use the spectrum of primordial gravity waves to further constrain the RH side of~(\ref{ub}). The initial spectrum of gravitational waves produced during inflation depends on $r$ and the tensor spectral index $n_{\rm T}$. However, their subsequent evolution depends on the post-inflationary thermal history and is in particular affected by an epoch of EMD~\cite{Durrer,Moroi} (also, see~\cite{DS,BH,F}). Tensor modes that enter the horizon during EMD, and the modes that are already at subhorizon scales, experience a suppression, compared to a standard thermal history, due to entropy generation in this epoch. As a result, the shape of the tensor spectrum is sensitive to the beginning and the end of the EMD phase, equivalently $T_{\rm O}$ and $T_{\rm R}$. This is complementary to the information from the scalar spectral index, mentioned above, which only depends on the duration of the EMD period encoded in $T_{\rm O}/T_{\rm R}$. Therefore, if $r$ is not much smaller than the current experimental bound $r_{\rm max} \simeq 0.064$~\cite{Planck18}, a future detection by (or limits from) the gravitational wave detectors could further constrain the allowed regions in the $T_{\rm O}-T_{\rm R}$ plane.            

It is known that DM perturbations at very small scales can grow during an epoch of EMD under suitable circumstances leading to formation of DM microhalos and a boost in the DM annihilation signal~\cite{Erickcek}. As shown in~\cite{Blanco}, the boost factor can reach values as high as $10^{18}$ in some cases. Taking this boost into account, it is possible to obtain upper bounds on $\langle \sigma_{\rm ann} v \rangle_{\rm f}$ as small as $10^{-32}$ cm$^3$ s$^{-1}$ from current gamma-ray observations~\cite{Linden}. However, cases with $\langle \sigma_{\rm ann} v \rangle_{\rm f} \ll 10^{-32}$ cm$^3$ s$^{-1}$ will be probably out of reach of indirect detection experiments in the near future. For the same reason, one can expect that this regime will likely escape direct detection as well. For very small values of $\langle \sigma_{\rm ann} v \rangle_{\rm f}$, with the help of~(\ref{ub}), a combination of collider experiments and cosmological observations could be used as an indirect test of an otherwise elusive scenario.

\subsection{An Example}

We now discuss a specific particle physics scenario of EMD to see how the inequality in Eq.~(\ref{ub}) is translated into constraints on the underlying model parameters. As mentioned before, moduli that arise in string theory are natural candidates for driving a period of EMD~\cite{KSW}. Consider a modulus field $\phi$ with mass $m_\phi$. It has gravitationally suppressed coupling to other fields resulting in a decay width:
\be \label{moddec}
\Gamma_\phi = {c \over 2 \pi} {m^3_\phi \over M^2_{\rm P}},
\ee
where typically $c \sim {\cal O}(0.1)$. 

\begin{figure}[h!]
  \centering
  \includegraphics[width=0.5\textwidth]{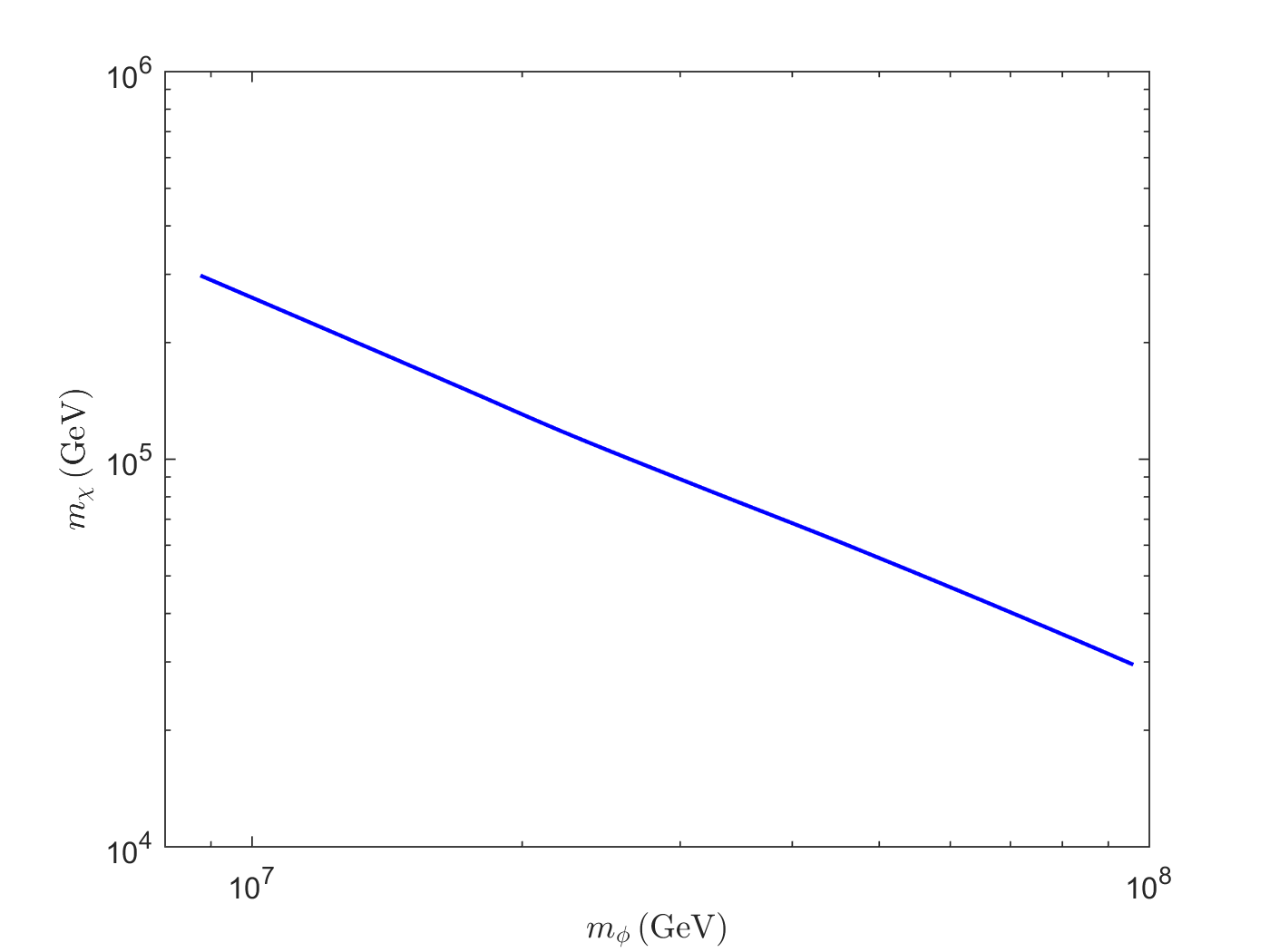}
  \captionsetup{justification = raggedright}
  \caption{Upper bound on the DM mass \(m_\chi\) for a given modulus mass \(m_\phi\) in the case of modulus-driven EMD. The curve is obtained numerically from the early-equilibrium regime. }
  \label{fig:7}
\end{figure}

The modulus $\phi$ gets displaced from the minimum of its potential during inflation, and starts oscillating about it when $H \simeq m_\phi$. Generic arguments based on effective field theory estimates~\cite{Dine,Holman,Linde,Dine2}, or explicit calculations~\cite{CDMQ}, suggest that the initial amplitude of oscillations is $\phi_0 \gtrsim {\cal O}(0.1 M_{\rm P})$. This implies that $\phi$ oscillations, which behave like matter, dominate the universe after their start, hence $H_{\rm O} \lesssim m_\phi$. Oscillations eventually decay when $H \simeq \Gamma_\phi$ and establish a RD universe with temperature $T_{\rm R}$~\footnote{Explicit examples in the context of flux compactifications where $\phi$ is the volume modulus are discussed in~\cite{ADS2,ACDS1}.}. Since the universe is approximately RD at the onset and the end of the EMD phase, and after using Eq.~(\ref{moddec}) with $c \sim 0.1$, the inequality in~(\ref{ub}) can now be written as: 
\be \label{modub}
m_\chi  \lesssim 120 \; g_{*,{\rm dec}} \left({g_{*,{\rm R}} \over g_{*,{\rm O}}}\right)^{1/4} \left({M_{\rm P} \over 10^{10} m_\phi}\right) (1 ~ {\rm GeV}),
\ee
where the RH side is basically controlled by $m_\phi$.

In Fig.~\ref{fig:7}, we show the allowed region of the $m_\chi-m_\phi$ plane accordingly. The curve depicts the upper bound of~(\ref{modub}), obtained from full numerical calculations. We have taken $g_{*,{\rm O}} = 106.75$ corresponding  to the high values of $T_{\rm O}$. Including new degrees of freedom beyond the SM (for example $g_{*,{\rm O}} = 228.75$, as in its minimal supersymmetric extension) will shift the curve slightly. Note that the slope of the curve does change, as expected, due to the implicit dependence of $g_{*,{\rm R}}$ on $m_\phi$ when $T_{\rm R}$ drops below the electroweak scale.

\section{Discussion and Conclusion}

We now turn to discussing how our results may be extended to more general situations that involve a non-standard thermal history.

Throughout the paper, we have considered a non-standard thermal history where the EMD epoch is preceded by a RD phase established at the end of inflationary reheating. However, for a very slowly decaying inflaton, the field(s) driving EMD may have comparable energy density to the inflaton before reheating completes. An example is a modulus-driven EMD scenario with $H_{\rm MAX} \ll m_\phi$~\footnote{For an explicit model, see~\cite{ACM}.}. In this case, the universe does not enter a truly RD phase at the end of inflationary reheating but rather a phase where the energy density of the radiation and matter components are roughly equal. One may approximate this case by taking $T_{\rm MAX} = T_{\rm O}$ in the non-standard thermal history considered here. We have checked that full numerical calculations are in very good agreement with this approximation. Such a scenario is also of the ``two-field" type, discussed in \cite{AO}, where the two matter-components have equal initial energy densities. We note that the freeze-in and freeze-out sides of the curves in Figs.~\ref{fig:2}-\ref{fig:5} are thus expected to merge at very high DM masses, $m_\chi > T_{\rm O}$, mimicking the shape of the baseline curve, though offset to smaller $\langle \sigma_{\rm ann} v \rangle_{\rm f}$. 

A general post-inflationary thermal history may include multiple epochs of EMD with respective parameters $T_{{\rm O},i}$ and $T_{{\rm R},i}$, $1 \leq i \leq n$, separated by intermediate RD phases. Because of the temperature dependence discussed in the Appendix, the EMD component of the DM relic abundance is mainly due to production in the last (i.e., $n$-th) bout of EMD. This implies that $T_{\rm R}$ must be replaced by $T_{{\rm R},n}$ in Eq.~(\ref{EMDdens}). The pre-EMD component, in the decoupling regime, is dominated by production at temperature $T_{\rm MAX}$, while in the early-equilibrium regime, it is set at temperature $T_{\rm dec}$ when chemical decoupling of DM particles occurs. Hence, the main modification to the expressions in~(\ref{preEMDdens}) is replacing $T_{\rm R}/T_{\rm O}$ with the product of $T_{{\rm R},i}/T_{{\rm O},i}$ to take all (relevant) EMD periods into account.

Throughout this paper, we have assumed $\langle \sigma_{\rm ann} v \rangle_{\rm f}$ to be constant within the temperature range of interest. In general, however, $\langle \sigma_{\rm ann} v \rangle_{\rm f}$ can have a temperature dependence, $\langle \sigma_{\rm ann} v \rangle_{\rm f}\propto T^n$. As discussed in the Appendix, early production of DM will remain to be important for a wide range of DM masses for $n \geq -1$. An example of such a case
is $\langle \sigma_{\rm ann} v \rangle_{\rm f} \propto T^2$, which happens when DM interacts with particles in the thermal bath through dimension-6 operators involving four fermions (similar to interaction of neutrinos with the thermal bath via $Z$ and $W$ exchange). Examples with larger values of $n$ include models studied in~\cite{Garcia}. In these cases, the allowed $\langle \sigma_{\rm ann} v \rangle_{\rm f}$ in Figs.~\ref{fig:2}-\ref{fig:5} should be interpreted as their values at the relevant temperature (i.e., $T \sim m_\chi/4$ in region 1, $T \simeq T_{\rm dec}$ in region 2, and $T \simeq T_{\rm MAX}$ in region 3). For $n > 0$, this implies that the pre-EMD production of DM can dominate even when $\langle \sigma_{\rm ann} v \rangle_{\rm f}$ at energies of ${\cal O} (m_\chi)$ is much smaller than the values shown in Figs.~\ref{fig:2}-\ref{fig:5}. On the other hand, as mentioned in the Appendix, DM production in the early RD phase and the memory phase of EMD will not be as important when $n < -3/2$. This is because the peak DM contribution occurs at $T \sim m_\chi$ for these cases. An important example of this case is the feebly interacting massive particle (FIMP) scenario~\cite{FIMP} where DM interacts with the thermal bath via mediators whose mass is below $m_\chi$ resulting in $n = -2$.

In conclusion, early stages in non-standard thermal histories that include a period of EMD can significantly affect production of DM via freeze-in for weak scale DM masses. We have demonstrated this in a post-inflationary history involving an early RD phase followed by an EMD epoch that reheats the universe to a temperature $T_{\rm R}$ before BBN. In the case where $\langle \sigma_{\rm ann} v \rangle_{\rm f}$ is constant over the temperature range of interest, the pre-EMD component of the DM relic abundance depends on the temperature at the onset of the EMD phase $T_{\rm O}$ and the reheating temperature after inflation $T_{\rm MAX}$, in addition to $T_{\rm R}$. This opens up vast regions of the $m_\chi-\langle \sigma_{\rm ann} v \rangle_{\rm f}$ plane where pre-EMD production can totally dominate the relic abundance as shown in Figs.~\ref{fig:2}-\ref{fig:5}. 

Moreover, DM particles reach chemical equilibrium in the pre-EMD era for a very broad range of $\langle \sigma_{\rm ann} v \rangle_{\rm f}$ spanning over many decades. The relic abundance in this case is virtually independent of $\langle \sigma_{\rm ann} v \rangle_{\rm f}$, and avoiding DM overproduction yields an inequality between $m_\chi$ and $T_{\rm O}/T_{\rm R}$. This brings an interesting possibility of combining collider searches (to measure $m_\chi$) with CMB and gravitational wave detector experiments (to constrain $T_{\rm O}/T_{\rm R}$) to test an elusive scenario that escapes indirect and direct detection.
We leave a detailed study of this for a future investigation.

\vskip 1.1cm
\section*{Acknowledgements}

This work is supported in part by NSF Grant No. PHY-1720174. We wish to thank the Campus Observatory at the University of New Mexico for providing computing time.

\vskip 1cm
\section{Appendix}

\subsection{Models with Constant $\langle \sigma_{\rm ann} v \rangle_{\rm f}$}

As an explicit example, we consider the model originally proposed in~\cite{Rabi}. This is an extension of the minimal supersymmetric standard model (MSSM) that includes new iso-singlet color-triplet superfields ${\bf X}$ and ${\bf {\bar X}}$ with respective hypercharges $+4/3$ and $-4/3$, and a singlet superfield ${\bf N}$ with the following superpotential:
\begin{eqnarray} \label{superpot}
W = W_{\rm MSSM} +  \lambda_{i} {\bf X} {\bf N} {\bf u^c_i} + \lambda^\prime_{i j} {\bf {\bar X}} {\bf d^c_i} {\bf d^c_j} + M_X {\bf X} {\bf {\bar X}} + \frac{M_N}{2} {\bf N} {\bf N} \, .
\end{eqnarray}
Here, ${\bf u^c}$ and ${\bf d^c}$ are superfields that contain the right-handed (RH) up-type and down-type quarks respectively, and $i$ and $j$ denote flavor indices (color indices are omitted for simplicity) with $\lambda^\prime_{i j}$ being antisymmetric under $i \leftrightarrow j$. The fermionic component of ${\bf N}$ and the scalar components of ${\bf X},~{\bf {\bar X}}$ are assigned to be even under $R$-parity. 

This model can accommodate new DM candidates in addition to those in MSSM. The scalar component of ${\bf N}$, denoted by ${\tilde N}$, is one candidate provided that it is the lightest supersymmetric particle (LSP)~\cite{Rabi,ADMS}. The fermionic component of ${\bf N}$ (denoted by $N$) is also a DM candidate if it is almost degenerate in mass with the proton $M_N \approx m_p$~\cite{AD}.  

At energies below $M_X$, the ${\bf X}$ and ${\bf {\bar X}}$ superfields can be integrated out to yield an effective $n=4$ superpotential term $\lambda \lambda^{\prime} {\bf N} {\bf u^c} {\bf d^c} {\bf d^c}/M_X$. Production of both DM candidates (${\tilde N}$ and $N$) from particles in the thermal bath proceeds through the corresponding dimension-5 operators at the component field level\footnote{We note that ${\tilde N}$ DM is also produced from decay of the next to lightest supersymmetric particle (NLSP). However, this contribution will be subdominant if the NLSP is underproduced during EMD (which can happen, for example, if its annihilation rate is above $3 \times 10^{-26}$ cm$^3$ s$^{-1}$). In the case of $N$ DM, a stable LSP provides a second component of DM. Again, LSP underproduction during EMD will render this component negligible. One can remove this component altogether by abandoning $R$-parity conservation as stability of $N$ is tied to stability of the proton when $M_N \approx m_p$~\cite{AD}.}. For temperatures $M_{\rm DM} < T < M_X$, this results in $\langle \sigma_{\rm ann} v \rangle_{\rm f} \propto \vert \lambda \lambda^{\prime}\vert^2/M^2_X$.

Another possible example where DM interacts with the thermal bath via a dimension-5 operator is in the context of the nonminimal supersymmetric standard model (NMSSM) with the following superpotential\footnote{One can also write a slightly different superpotential $W \supset \lambda {\bf \Phi} {\bf H_u} {\bf H_d} + y {\bf {\bar \Phi}} {\bf S} {\bf S} + M_\Phi {\bf {\bar \Phi}} {\bf \Phi}$ based on the NMSSM model in~\cite{Murayama}. This model includes two singlet Higgs superfields, ${\bf \Phi}$ and ${\bf {\bar \Phi}}$, instead of one. Integrating out the ${\bf \Phi}$ and ${\bf {\bar \Phi}}$ yields to the same $n=4$ superpotential term as integrating out ${\bf {\Phi}}$ in~(\ref{nmssm}) does.}:
\begin{eqnarray} \label{nmssm}
W & = & W_{\rm MSSM} + \lambda {\bf \Phi} {\bf H_u} {\bf H_d} + y {\bf \Phi} {\bf S} {\bf S} + {M_\Phi \over 2} {\bf \Phi} {\bf \Phi} + {M_S \over 2} {\bf S} {\bf S} \, .
\end{eqnarray}
Here, ${\bf H_u}$ and ${\bf H_d}$ are the superfields containing the MSSM Higgs fields, and ${\bf \Phi}$ and ${\bf S}$ are singlet superfields whose scalar components are even under $R$-parity. The fermionic component of ${\bf S}$, denoted by ${\tilde S}$, can be the DM candidate in this model if it is the LSP. At energies below $M_\Phi$, the ${\bf \Phi}$ superfield can be integrated out to yield the $n=4$ superpotential term $y \lambda {\bf H_u} {\bf H_d} {\bf S} {\bf S}/M_\Phi$. Production of DM candidate ${\tilde S}$ then proceeds through a dimension-5 operator involving two Higgs fields, which results in $\langle \sigma_{\rm ann} v \rangle_{\rm f} \propto \vert y \lambda \vert^2/M^2_\Phi$ for $M_S < T < M_\Phi$.

\subsection{Calculation of DM Relic Abundance}

Here, we present the details of calculating the DM abundance produced prior to EMD as well as in the memory phase of EMD.

We will begin with with the late phase of EMD, i.e., $H \ll H_{\rm tran}$~(see (\ref{htran1})). In the freeze-in scenario, the rate for production of DM particles from the annihilation of SM particles, $\Gamma_\chi = \langle \sigma_{\rm ann} v \rangle_{\rm f}~  n_{\chi,{\rm eq}}$, is, by definition, small compared to the Hubble rate $H$. At sufficiently high temperatures, $T \gg m_\chi$, we have $n_{\chi,{\rm eq}} \propto T^3$. Assuming that $\langle \sigma_{\rm ann} v \rangle_{\rm f}$ is constant, as mentioned before, and using the expression in Eq.~(\ref{tinstEMD}), we see that $\Gamma_\chi \propto H^{3/4}$. This implies that the freeze-in condition is satisfied more strongly at earlier times (equivalently higher temperatures). Then, since $n_{\chi,{\rm eq}}$ and $\Gamma_\chi$ are Boltzmann suppressed at $T \ll m_\chi$, DM particles will not reach chemical equilibrium during EMD provided that $\Gamma_\chi \ll H$ when $T \sim m_\chi$.

The situation, however, is different prior to the entropy generating phase of EMD (i.e., $H \gtrsim H_{\rm tran}$). In the RD phase after inflationary reheating, see Eq.~(\ref{tinstrad}), $T \propto H^{1/2}$ implying that $\Gamma_\chi \propto H^{3/2}$, while, during the memory phase of EMD, see Eq.~(\ref{tinsttran}), $T \propto H^{2/3}$ and hence $\Gamma_\chi \propto H^2$. As a result, DM production becomes more efficient at earlier times (higher temperatures) for $H \gtrsim H_{\rm tran}$. Then, in order for DM to not be in chemical equilibrium with the thermal bath prior to the entropy generating EMD epoch, we need $\Gamma_\chi \ll H$ at $H \simeq H_{\rm MAX}$. Based on this, DM production at $H_{\rm tran} \lesssim H \lesssim H_{\rm MAX}$ has two regimes, which we discuss separately below.      
\vskip 5mm
\noindent
{\bf Decoupling regime.} Production of DM from SM particles in the thermal bath will be inefficient, and DM will never reach chemical equilibrium in the post-inflationary universe, if $\Gamma_\chi \ll H$ when $H \simeq H_{\rm MAX}$. This is the case if:
\be \label{ficond}
\langle \sigma_{\rm ann} v \rangle_{\rm f} \ll {\pi^3 g^{1/2}_{*,{\rm MAX}} \over \sqrt{90} \zeta(3) M_{\rm P} T_{\rm MAX}}.
\ee
In this regime, $n_\chi \ll n_{\chi,{\rm eq}}$ and the third equation of Eq.~(\ref{boltzmann}) results in:  
\begin{equation} \label{comoving1}
\frac{d(a^3n_\chi)}{dt} \approx a^3\langle\sigma_{\rm ann} v\rangle_{\rm f} ~ n_{\chi,{\rm eq}}^2.
\end{equation}
Integrating both sides between $H_{\rm tran}$ and $H_{\rm MAX}$, and after converting $dt$ to $dH$, the comoving number density of DM is found to be: 
\be \label{comoving2}
(a^3n_\chi)_{\rm dec} \simeq {\zeta(3)^2 \over \pi^4} \langle \sigma_{\rm ann} v \rangle_{\rm f} \left(
\int_{H_{\rm tran}}^{H_{\rm O}}{{2 T^6 a^3 \over 3 H^2} dH} +  \int_{H_{\rm O}}^{H_{\rm MAX}}{{T^6 a^3 \over 2 H^2} dH} \right).
\ee
Here, we have used $t=2/3H$ for $H_{\rm R} \lesssim H \lesssim H_{\rm O}$, $t = 1/2H$ for $H_{\rm O} \lesssim H \lesssim H_{\rm MAX}$, and $n_{\chi,{\rm eq}} = \zeta(3) T^3/\pi^2$ as $T \gg m_\chi$ for much of the pre-EMD phases and $n_{\chi,{\rm eq}}$ is Boltzmann suppressed for lower temperatures. We have taken $\langle \sigma_{\rm ann} v \rangle_{\rm f}$ out of the integrals as it is assumed to be a constant. 

Using the relation in Eq.~(\ref{tinstrad}) and $a \propto H^{-1/2}$ for $H_{\rm O} \lesssim H \lesssim H_{\rm MAX}$, as well as the relation in Eq.~(\ref{tinsttran}) and $a \propto H^{-2/3}$ for $H_{\rm R} \lesssim H \lesssim H_{\rm O}$, we see that both integrals on the RH side of~(\ref{comoving2}) are dominated by their upper limits. Thus, for $H_{\rm MAX} \!\gg\! H_{\rm O}$, production prior to EMD dominates over the memory phase.\!
%
%
We are interested in $n_\chi$ normalized by the entropy density $s = 2 \pi^2 g_* T^3/45$ once significant entropy production has stopped, where $T = T_{\rm R}$, which results in:
\be \label{fifinal}
\left({n_\chi \over s} \right)_{\rm dec} \simeq {45 \sqrt{90} \zeta(3)^2 \over 2\pi^7 g^{5/4}_{*,{\rm MAX}} g_{*,{\rm O}}^{1/4}} ~ \langle \sigma_{\rm ann} v \rangle_{\rm f} M_{\rm P} ~ \left({T_{\rm MAX} T_{\rm R} \over T_{\rm O}}\right) .
\ee
\vskip 2mm
\noindent
{\bf Early-equilibrium regime.} If $\Gamma_\chi \gtrsim H$ when $H \simeq H_{\rm MAX}$, then DM particles will initially be in chemical equilibrium with the thermal bath in the post-inflationary universe. This is satisfied if:
\be \label{deccond}
\langle \sigma_{\rm ann} v \rangle_{\rm f} \gtrsim {\pi^3 g^{1/2}_{*,{\rm MAX}} \over \sqrt{90} \zeta(3) M_{\rm P} T_{\rm MAX}}.
\ee
However, as mentioned before, $\Gamma_\chi$ decreases more quickly with time than $H$ in the prior RD and memory periods. This implies that DM can drop out of chemical equilibrium sometime prior to the late EMD phase and stay so during the rest of the EMD epoch. DM chemical decoupling occurs at a temperature $T_{\rm dec}$, where $T_{\rm tran} \lesssim T_{\rm dec} \lesssim T_{\rm MAX}$, at which $\Gamma_\chi$ drops below $H$. We note that for  very high DM masses, such that $T_{\rm tran} \ll m_\chi$, decoupling would have to happen before $T \sim m_\chi$ in order to avoid transitioning to a freeze-out regime. 

The number density of DM particles follows its equilibrium value $n_{\chi,{\rm eq}} \propto T^3$ down to $T_{\rm dec}$. At lower temperatures, $n_\chi$ is redshifted $\propto a^{-3}$ due to expansion of the universe and the comoving number density of DM remains essentially constant.
%
%
Since production of radiation from the decaying component(s) driving EMD is negligible for $H_{\rm tran} \lesssim H \lesssim H_{\rm MAX}$, the comoving entropy density is constant implying that $n_\chi \propto g_* T^3$ in this interval. We then find:
\be \label{dec2}
(a^3 n_\chi)_{\rm e\text{-}eq} \simeq {\zeta(3) g_{*,{\rm O}} \over \pi^2 g_{*,{\rm dec}}} T^3_{\rm O} a^3_{\rm O}.
\ee
Eventually, after normalizing $n_\chi$ by $s$ at the end of EMD where $T = T_{\rm R}$, we arrive at:
\be \label{decfinal}
\left({n_\chi \over s}\right)_{\rm e\text{-}eq} \simeq {45 \zeta(3) \over 2 \pi^4 g_{*,{\rm dec}}} ~ \left({T_{\rm R} \over T_{\rm O}}\right) .
\ee
\vskip 2mm
An important point to keep in mind is that in both of the decoupling and early equilibrium regimes, Eqs.~(\ref{fifinal}) and~(\ref{decfinal}) respectively, the relic abundance is set by DM production in the pre-EMD phase. In the decoupling regime, production at $T \simeq T_{\rm MAX}$ makes the most important contribution. In the early equilibrium regime, DM particles reach chemical equilibrium in the pre-EMD era and their comoving number density remains constant after they decouple.
\vskip 2mm
Finally, we comment on how early production of DM changes in more general cases when the annihilation rate depends on temperature, $\langle \sigma_{\rm ann} v \rangle_{\rm f} \propto T^n$. By keeping $\langle \sigma_{\rm ann} v \rangle_{\rm f}$ inside the intergrals in Eq.~(\ref{comoving2}), we see that the first integral is dominated by its upper limit for $n \geq -3/2$, while the same happens for the second integral if $n \geq -1$. Therefore, the importance of early production of DM and the appearance of the decoupling and early-equilibrium regimes will qualitatively remain unchanged as long as $n \geq -1$. Quantitatively, our results will become stronger for $n > 0$ and weaker for $-1 \leq n < 0$.



\end{document}